# FERROFLUID BASED DEFORMABLE MIRRORS - A NEW APPROACH TO ADAPTIVE OPTICS USING LIQUID MIRRORS

P. Laird[a], R. Bergamasco[b], V. Bérubé[a], E.F. Borra[a], A. Ritcey[b], M. Rioux[a], N. Robitaille[a,c],
S. Thibault[c], L. Vieira da Silva Jr.[a], H. Yockell-Lelièvre[b]

[a]Laval University, Dept. of Physics/Centre d'optique, photonique et lasers; [b]Laval University, Dept. of Chemistry, [c]INO, Sainte-Foy (Quebec, Canada)

## ABSTRACT

The trend towards ever larger telescopes and more advanced adaptive optics systems is driving the need for deformable mirrors with a large number of low cost actuators. Liquid mirrors have long been recognized a potential low cost alternative to conventional solid mirrors. By using a water or oil based ferrofluid we are able to benefit from a stronger magnetic response than is found in magnetic liquid metal amalgams and avoid the difficulty of passing a uniform current through a liquid. Depositing a thin silver colloid known as a metal liquid-like film (MELLF) on the ferrofluid surface solves the problem of low reflectivity of pure ferrofluids. This combination provides a liquid optical surface that can be precisely shaped in a magnetic field. We present experimental results obtained with a prototype deformable liquid mirror based on this combination.

## 1. INTRODUCTION

The surface of a liquid follows an equipotential surface to a very high precision so that one can have desired characteristics by shaping the equipotential surfaces with force fields. For example, using gravity alone it is possible to make parabolic mirrors by rotating a container covered by a reflective liquid around a vertical axis of rotation. Laboratory work [1,2] has shown that rotating liquid mirrors have very good surface qualities. Liquid Mirror Telescopes (LMT) have been built and used to obtain astronomical data[3] as well as for the atmospheric sciences [4,5,6,7]. The main advantage of liquid mirrors is their low cost stemming from the lack of high-precision grinding and polishing steps in their fabrication. Clearly it would be useful to have liquid mirrors with shapes other than a parabola. Using externally applied magnetic fields to modify the equipotential due to gravity is a way of achieving this goal.

Magnetically shaped liquid mirrors have been proposed in the scientific literature. Whitehead and Shutter[8] proposed using an iron-mercury amalgam to transform the parabolic shape resulting from a rotating liquid into a sphere. Ragazzoni and Marchetti[9] proposed to deform mercury with an electric current flowing in the liquid in an external magnetic field to make a surface useful for astronomical instrumentation. While mercury is ideally suited for conventional liquid mirrors because of its high reflectivity and low melting temperature, as a magnetic liquid it has problems. Firstly, the chemistry of mercury is not optimal if one wants to obtain a high magnetic susceptibility; specifically it is very difficult to obtain a stable metallic based magnetic liquid[10]. Second, the high density of mercury necessitates a larger deforming force and thus a strong magnetic field. As most concepts for generating the shaping field involve the use of magnetic coils, the increased ohmic heating produced by the larger currents creates a technical challenge to building a practical mercury based deformable mirror. For these reasons, it would be useful to make magnetically shaped reflective liquids having better physical and chemical characteristics.

In this paper, we describe a new type of reflective optical element made of a magnetic fluid surrounded by a metallic layer. The liquid in the middle is a ferrofluid: a superparamagnetic liquid that can be made using low density carrier fluids so that a relatively small magnetic field can be used to shape the liquid surface. As a consequence, one can impart any desired shape to the reflecting surface by introducing into the liquid the appropriate magnetic field geometry[11]. The principal application discussed in this paper is a deformable mirror (DM) for use in adaptive optics (AO). Proposed extremely large telescopes (ELT) projects as well as so called Extreme Adaptive Optics (ExAO) are driving the demand for larger numbers of actuators in the DM. Current technologies become very costly at high actuator counts. Ferrofluid based systems offer a potentially low cost solution for these applications.

## 2. METAL LIKE LIQUID FILMS (MELLFS)

The air-oil or air-water interface of an uncoated magnetic liquid exhibits inherently low reflectivity. Highly reflecting liquid surfaces can, however, be obtained by the application of a thin film composed of silver nanoparticles. Stable interfacial suspensions of silver particles have been described in the literature[12] and are frequently referred to as Metal Liquid-Like Films, or MELLFs. These intriguing systems combine the optical properties of metals with the fluidity of a liquid suspension and are therefore well adapted to applications in the field of liquid optics. The MELLF forms an extremely thin layer that follows the substrate very closely, allowing precise control of the reflective surface. The fabrication of a MELLF involves the creation of silver nanoparticles, generally by chemical reduction of a silver salt in aqueous solution, and the subsequent coating of the particles with an organic ligand. When coated, the particles are no longer stable in the aqueous phase and spontaneously assemble at the water-organic interface. The role of the surfactant is paramount to both the surface assembly of the particles and their stabilisation against aggregation. Despite a systematic study involving a large variety of ligands[13], the detailed surface chemistry and the precise mechanism of the formation of a MELLF remains unclear.

Similar interfacial films using gold have also been demonstrated and other metals may also be used to tailor the reflectivity and spectral response of the resulting surface to the application. In a recent article[14], we showed preliminary results obtained with an interfacial MELLF. For that MELLF, the colloidal particles were trapped at the interface between an aqueous solution and a denser organic liquid. We have improved the technology so that we can now coat the top surface of the liquid and can coat oils. Coating oils is a major improvement for we now can use a much greater variety of liquids. Oils can support a greater concentration of magnetic particles giving them a higher magnetic susceptibility with a corresponding decrease in the power requirements for a given deformation. Furthermore, the use of oils solves the evaporation problem that occurred with water.

## 3. FERROFLUIDS

Magnetic liquids were sought after for many years for a wide range of applications. Initial attempts to use molten ferromagnetic materials failed because the magnetic properties change drastically at the Curie temperature, and there are no known materials for which the Curie temperature is higher than the melting point. The solution was found by using a multi-phase liquid known as a ferrofluid in which ferri- or ferromagnetic particles are held in a colloidal suspension in a carrier liquid. Magnetic particles have a strong tendency to agglomerate; in a stable ferrofluid agglomeration is prevented by the use of a surfactant. A wide range of carrier liquids is available and the choice of carrier has a very strong impact on the hydrodynamic properties of the resulting ferrofluid. In the presence of an external magnetic field, the magnetic particles will align themselves with the field and the bulk liquid becomes magnetized. The resulting behavior is similar to that of paramagnetic liquids such as liquid oxygen but many times stronger which leads to the use of the term superparamagnetism.

The mechanical behavior of a ferrofluid is governed by a combination of magnetic and hydrodynamic forces. In practical terms, the interplay of these forces determines both the magnitude and speed of deformations that can be achieved for a given external magnetic field. Stable ferrofluids can be produced that exhibit an extremely wide range of properties and there is often a compromise between magnetic susceptibility and hydrodynamic properties.

**Ferrofluid relaxation time**

Ferrofluids relax in two ways when an external magnetic field is removed; the first is hydrodynamic or Brownian motion[15] and is governed by the combined properties of the magnetic material and coating agent.

$$\tau_B = \frac{V\eta_0}{kT} \qquad (1)$$

Where V is the hydrodynamic particle volume, $\eta_0$ is the viscosity of the carrier liquid. This mechanism is also referred to as extrinsic relaxation because it relates to the entire particle. The relaxation second mechanism is Néel[16] or intrinsic relaxation; it involves magnetic domains and is basically independent of surface effects. The driving force in this mechanism is thermal perturbation of the particle magnetization.

$$\tau_N = \frac{1}{f_0} \exp\left(\frac{KV}{kT}\right) \qquad (2)$$

Where K is the anisotropy constant and $f_0$ is approximately $10^9$ Hz. Eqn 2 is valid for the condition $KV \ll kT$, that is to say when the energy barrier due to magnetic anisotropy is smaller than thermal energy available to induce fluctuations in the magnetism. By comparing equations 1 and 2 we see that Brownian relaxation dominates for large particle sizes. For kerosene based fluids with magnetite particles, the transition occurs around 10 nm at which point the relaxation time is on the order of $10^{-8}$ s. At the time scales of interest in AO, this fast relaxation time means the ferrofluid is effectively free of hysteresis.

**Deformation in a magnetic field**

As a first approximation, the deformation of an inviscid magnetic liquid in a static magnetic field normal to the surface of the liquid is governed by gravity and the magnetic body force on the liquid arising from an external magnetic field. This deformation was described by Jones[17]:

$$\Delta h = \frac{B^2(\mu_r - 1)}{2\mu_0 \rho g} \qquad (3)$$

Where $\rho$ is the density of the ferrofluid and B is the external magnetic field just above the ferrofluid surface. For optical deformations on the order of several microns, the required change in magnetic field strength is very small and the relative permeability $\mu_r$ can be assumed to be constant.

The thickness of a typical MELLF layer is approximately 150 nm and so can be neglected for this analysis. It is possible that some dilution of the ferrofluid may occur during the deposition of the reflective layer but this is easily accounted for by modifying the density and magnetic susceptibility terms in eqn 3.

When the normal component of the magnetic field exceeds a critical value related to the interfacial tension and the density of the ferrofluid, a non-linear surface phenomenon known as the Rosensweig instability develops. This instability appears as a series of spikes that are essentially amplified surface waves and it effectively rendering the surface useless as a mirror. For typical kerosene based ferrofluids the Rosensweig instability occurs around 84 Gauss[18]. From eqn 3 this gives a maximum surface deformation before the onset on the instability on the order of 1 mm. This is well above the requirement for DM stroke requirement for AO, which is on the order of 10 µm.

**Requirements for the Extremely Large Telescope (XLT)**

The Extremely large telescope (XLT)[19] is a Canadian proposal on the 20 to 30 meter scale which may eventually replace the existing Canada France Hawaii Telescope. To illustrate the need for low cost deformable elements, it is interesting to consider the number of actuators required for a 20m class telescope based on seeing conditions at the current CFHT site. The coherence length[20], $r_0$ in the visible and the near infrared is 20 cm (R band @ 650 nm) and 93 cm (K band @ 2.23 µm) respectively. The number of actuators needed to achieve diffraction-limited performance is $\sim (D/r_0)^2$ where D is the primary mirror diameter. So, for a 20 m diameter we will need 9070 actuators in the visible at 650 nm and 462 actuators in the near infrared at 2.23 µm.

All AO systems require a reference, which may be a sufficiently bright natural guide star, an artificial laser guide star or in very rare cases, the science object itself. Due to anisoplanatism, the reference must be located within a small angle of astronomical target being observed or the correction applied will not be representative of the atmosphere along the path to the target. In general only a fraction of the sky is well-suited to science observation; these limitations are more severe in the visible than the infrared. One approach to reduce this problem is the use of multi-conjugate adaptive optics (MCAO). MCAO uses multiple references to provide a more accurate reconstruction of atmospheric turbulence along multiple paths; in practice this requires several DMs thus further increasing the need for low cost components.

## 4. EXPERIMENTAL

The prototype mirror is 80 mm in diameter with an actuator spacing of 5.5 mm, the layout is as shown in fig 1. For this series of tests, only 31 of the positions in the base were filled. The actuators were 5 mm diameter coils with a soft-ferrite core used to concentrate the magnetic flux and thus reduce current requirements by at least two orders of magnitude. The actuator coils are relatively easy to fabricate and lend themselves to low cost mass production. Another possibility for actuator elements is the use of a mircofabricated coil array of the type demonstrated by Cugat et al[12]. Commands were generated using a desktop PC running MATLAB and NuDAQ 6216V analog output cards driving custom build amplifier circuits as shown in fig 2. The mirror surface was measured using a Zygo general purpose interferometer (GPI).

The mirror was characterized by examining the influence function which is the deformation measured at an actuator due to an applied current in an adjacent coil. These measurements were carried out for various liquid thicknesses.

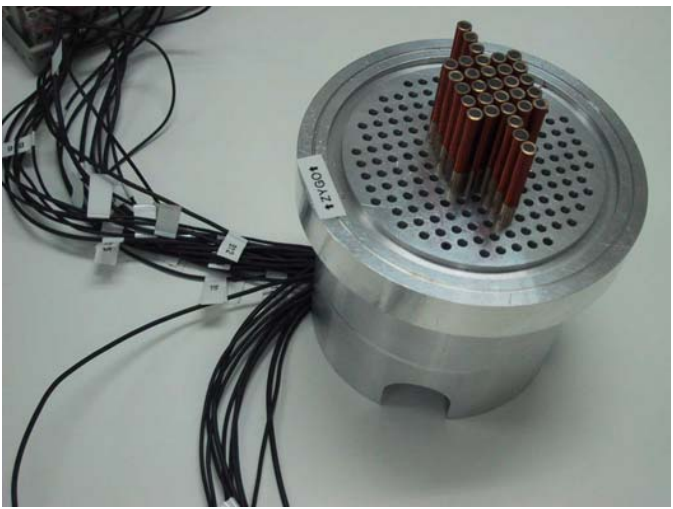
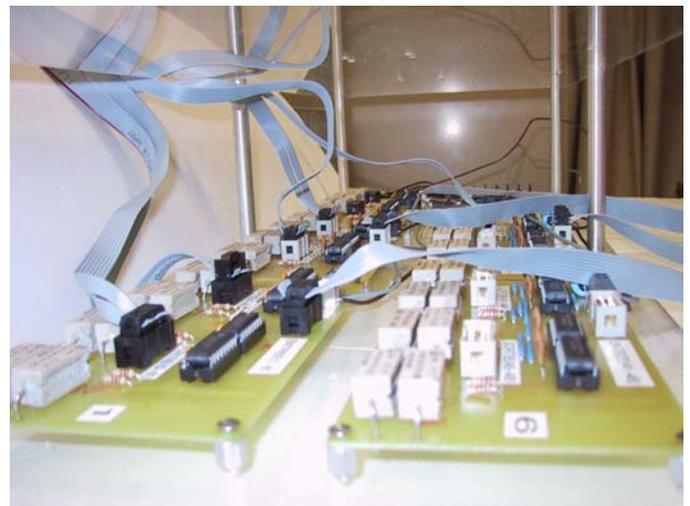

Figure 1: Mirror base without liquid container        Figure 2: Amplifier circuit boards

# 5. RESULTS

## 5.1 Deformation

The ultimate deformation limit of surface deformation is the onset of the Rosensweig instability; as discussed previously, this is not an issue for AO applications as the limit greatly exceeds the stroke requirement for a DM. A more practical limit arises from the heat dissipation in the actuator coils. Deformations of up to 10 μm have been measured with the GPI, beyond which point the measurement becomes difficult due to the extremely close fringe spacing.

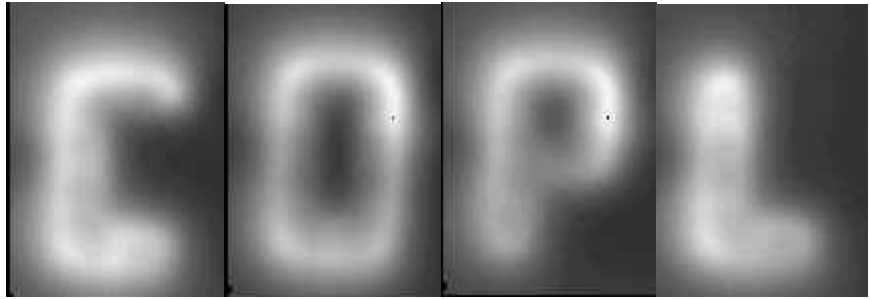

**Figure 3: Wavefront image for 5x5 uncoated ferrofluid based deformable mirror with 5.5 mm actuator spacing.**

Fig 3 shows a wavefront produced using the 25 central actuators in the mirror and measured with the GPI. Differences in magnetic field strength for a given current of to 10% were measured for uncalibrated coils. This result is in agreement with known manufacturing tolerances for the ferrite core material. The image in fig 3 is produced after conducting a manual calibration using a gaussmeter. Calibration using actual wavefront data is proving to be much more accurate.

The response time of these mirrors show a strong dependence on the size of the deformation and the viscosity of the carrier liquid. Efforts are currently underway to characterise these parameters more precisely. Early indications are that the mirror frequency will exceed 100 Hz for deformations on the order of 10 μm for typical kerosene based fluids.

## 5.2 Influence function and coupling

The inter-actuator coupling was measured for different mirror thicknesses and was found to vary between 18 and 33% as shown in Table 1. This result is comparable to existing, solid deformable mirrors. Coupling was observed to vary with ferrofluid depth, as is to be expected with the magnetic field lines becoming less parallel as the liquid surface is moved farther from the coils. For the range of deformations tested, the influence retains the same shape for increasing current. This result is encouraging because it implies that existing algorithms for controlling DMs can be adapted relatively easily to incorporate this type of liquid mirror. The symmetry of the influence function will reduce the need for involved computations in the control loop.

Table 1: Coupling as a function of ferrofluid thickness

| Thickness (mm) | Coupling (%) |
|---|---|
| 1.0 | 18 |
| 1.5 | 23 |
| 2.0 | 33 |

## 5.3 Reflectivity

Fig 4 shows the reflectivity curves for three different MELLF combinations. The best curve is for a MELLF on water and the other two curves are for MELLFs deposited on paraffin and on an oil-based ferrofluid respectively.

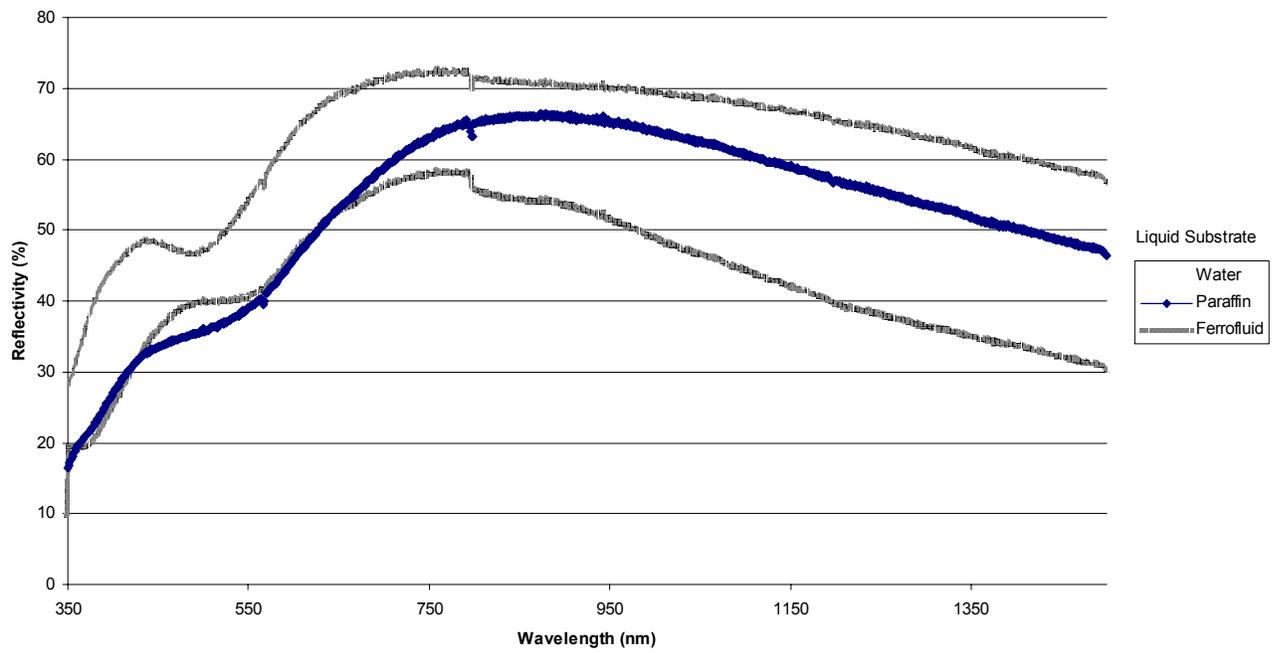

Figure 4: Liquid mirror reflectivity for different substrates

In the current mirror, chemical reactions between the MELLF and the ferrofluid reduce the reflectivity of the combined surface. The ferrofluid used is a comercially available product and the details of its chemical composition are proprietary, thus making it difficult to tailor the reflective layer to ensure chemical compatibility. We suspect that we eventually shall have to develop ferrofluids optimized for this particular application. The curve for the pure MELLF shows that a great deal of improvement is possible once this problem is addressed. We are currently carrying out work to improve the reflectivities of the MELLFs themselves. Control of parameters such as particle size distribution and packing fraction in the MELLF still have room for improvement and theoretical modelling has suggested several other changes in MELLF formulation that may lead to improved reflectivity. It is reasonable to conclude that we are still quite far from the limit of this technology.

### 5.4 Power requirements and cross-talk

Power dissipation in the actuators of a DM is a common problem because astronomical optics are often very temperature sensitive in order to preserve seeing quality. The use of high magnetic susceptibility ferrofluids with low density allows operation at low coil currents. A 10 μm deformation for a 2mm thick ferrofluid requires a power dissipation of approximately 8 mW per coil. At this level, no active cooling was required to use the DM at room temperature. Additionally, because the coils are low resistance and the currents required are quite small, no high voltage stage is required, unlike for piezoelectric actuators. Cross-talk in magnetic coil actuated mirrors due to mutual inductance in neighboring coils has been investigated by Cugat et al[21] for a similar actuator geometry and was found to be negligible. In our mirror cross-talk would be expected to appear as a secondary peak centered on an adjacent actuator, this was not observed and it can be concluded that the only influence mechanism between actuators can be described by a Gaussian profile.

## 6. CONCLUSION

We have demonstrated the concept of using magnetic liquids and reflective films to form a magnetically deformable liquid mirror capable of producing strokes of more than 10 μm with relatively low power requirements. The speed of this device for deformations up to 10 μm is expected to exceed 100 Hz and further testing is currently underway to confirm this. Deformations on the order of a millimetre are also possible using this system although the power requirements and heat loads become significant.

The influence function of a liquid deformable mirror has been characterised and found to have coupling in the range of 18-33% which is consistent with conventional deformable mirrors. From these results it can be concluded that control algorithms developed for conventional deformable mirrors can be easily adapted to the liquid mirror, thereby reducing the integration effort required to realise a complete system.

The next step is to complete 100 actuator prototype and integrate this mirror with a wavefront sensor to demonstrate correction of simulated aberrations. For larger mirrors we will need to move away from PC based analog output cards to digital signal processing boards due to the heavy computational burden associated with deformable mirror control. The integration of the amplifier hardware with the DSP boards is also being investigated. These results represent only one of a large number of possible combinations of magnetic liquids and reflective coatings and there is considerable room for improvement of the reflectivity characteristics in general as well as for specific wavelength bands.

There are many potential applications of this technology. Early indications are that these liquid mirrors can be used in an inverted configuration with the addition of a suitable biasing magnetic field. This may make it possible to integrate a liquid mirror as an adaptive secondary, located above the main mirror or in other interesting geometries.

## ACKNOWLEDGMENTS

This research was supported by the Natural Science and Engineering Research Council and The Canadian Institute for Photonics Innovation. L.Vieira da Silva Jr. wishes to thank to the Conselho Nacional de Desenvolvimento Científico e Tecnológico, CNPq , Brazil for their financial support.